\documentclass[prb,showpacs,twocolumn,amsmath,amssymb,superscriptaddress]{revtex4}
\usepackage{txfonts}
\usepackage{epsfig}
\usepackage{graphicx}
\usepackage{dcolumn}
\usepackage{bm}

\begin{document}
\title{Theoretical investigation of TbNi$_{5-x}$Cu$_x$ optical properties} 
\author{I.A.~Nekrasov}
\affiliation{Institute of Electrophysics, Amundsena Str. 106, 620016 Ekaterinburg,  Russia}
\author{E.E.~Kokorina}
\affiliation{Institute of Electrophysics, Amundsena Str. 106, 620016 Ekaterinburg,  Russia}
\author{V.A.~Galkin}
\affiliation{Institute of Electrophysics, Amundsena Str. 106, 620016 Ekaterinburg,  Russia}
\affiliation{Ural State University, Lenina pr. 51, 620083 Ekaterinburg, Russia}
\author{Y.I.~Kuz'min}
\affiliation{Institute of Metal Physics, S. Kovalevskoj Str. 18, 620219 Ekaterinburg, Russia}
\author{Y.V.~Knyazev}
\affiliation{Institute of Metal Physics, S. Kovalevskoj Str. 18, 620219 Ekaterinburg, Russia}
\author{A.G.~Kuchin}
\affiliation{Institute of Metal Physics, S. Kovalevskoj Str. 18, 620219 Ekaterinburg, Russia}

\begin{abstract}

In this paper we present theoretical investigation of optical conductivity for
intermetallic TbNi$_{5-x}$Cu$_x$ series. In the frame of LSDA+U
calculations electronic structure for $x=$ 0, 1, 2 and on top of that optical
conductivities were calculated. Disorder effects of Ni for Cu substitution on a level of LSDA+U
densities of states (DOS) were taken into account via averaging over all possible Cu ion
positions for given doping level $x$. Gradual suppression and loosing of structure of optical conductivity at 2 eV
together with simultaneous intensity growth at 4 eV correspond to increase of Cu and decrease
of Ni content. As reported before [Knyazev $et~al.$, Optics and Spectroscopy {\bf 104}, 360 (2008)]
plasma frequency has non monotonic doping behavior
with maximum at $x=1$. This behavior is explained as competition between
lowering of total density of states on the Fermi level $N(E_F)$ and growing of number of carriers.
Our theoretical results agree well with variety of recent experiments.
\end{abstract}

\maketitle

\section{Introduction}

Intermetallic compounds of RNi$_5$-type have been attracting
considerable interest initiated by their remarkable physicochemical properties and potential
applications as materials with high hydrogen storage capacity~\cite{france}. Their physical
properties, which are characterized by a large variety of magnetic structures and electronic
peculiarities, have been actively investigated, both theoretically and experimentally.
Substitution of one rare-earth metal R with another one modifies significantly these
characteristics because of crystal field effects and changes of exchange interaction value
between the conduction and 4f-localized electrons.

On the the other hand substitution of Ni for some other transition metal also leads to
interesting physical phenomena. For example, series of isostructural RNi$_{5-x}$Cu$_x$ alloys
was found experimentally to have non monotonic concentration dependence of the magnetic
susceptibility, Curie temperature T$_C$, electronic specific heat and
resistivity~\cite{Kuchin,Gurevich,Grechnev,Kuchin2006,Pirogov09}. A number of studies indicate a direct
correlation between this anomalous behavior of the parameters and band structure evolution with
copper content increase~\cite{Gurevich, Grechnev, Burzo}. To explain these experimental data
one needs more detailed investigation of electronic structure of the RNi$_{5-x}$Cu$_x$ series
for different $x$.

There are several experimental and theoretical works on the RNiCu compounds. X-ray
photoemission study of LaNi$_{5-x}$Cu$_x$ solid solution shows presence of Cu-3d band, which is
located almost 2 eV below the 3d band of Ni and is weakly hybridized with the latter
one~\cite{Burzo}. The calculation of Ref.~\onlinecite{Grechnev} predicts the existence of a
wide ($\sim$ 1,5 eV) peak in the electron density of states of the YNi$_{5-x}$Cu$_x$ alloys at
energies 2.5-4 eV below Fermi level E$_F$, which is related to the 3d electrons of Cu. The
evolution of frequency dependencies of the optical conductivity of the TbNi$_{5-x}$Cu$_x$ compounds upon
substitution of Ni for Cu atoms investigated in Ref.~\onlinecite{Knyazev08}. There it was
observed that a new structure arises in the optical spectra (a broad absorption maximum peaking
at 4 eV) while the Cu content increases. Detailed investigation of
crystal structure and thermodynamic properties from both experimental and band structure calculation
sides is done in Ref.~\onlinecite{Lizarraga06} (except for the optical properties).

Within this study we continue investigation of the electronic structure of TbNi$_{5-x}$Cu$_x$
isostructural alloy series with $x=$ 0, 1, 2 by carrying out self-consistent {\it ab~initio} LSDA+U
calculations. Disorder effects are taken into account in a combinatorial way with respect to
possible number of Cu positions for given doping. Corresponding averaged over the Cu positions
spin resolved densities of states are calculated. Following it interband contribution to the
theoretical optical conductivity is determined. Drude (intraband) contribution is taken as
a Lorentzian with experimentally determined parameters. Comparison of our theoretical results with
recent optical experiment is presented.

\section{Calculation of the electronic structure}

\begin{figure}[t] 
\epsfxsize=7cm \epsfbox{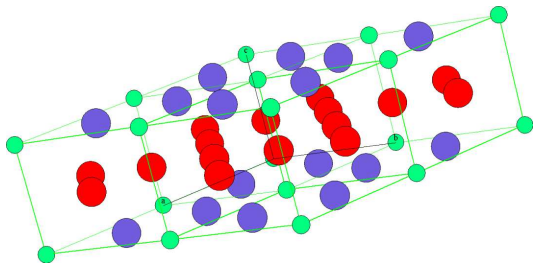}
\caption{(Color online)  TbNi$_5$ crystal structure. Tb ions -- green circles, Ni1,Ni2 --
violet and red circles correspondingly.}
\label{fig1}
\end{figure}

The TbNi$_5$ compound crystallizes into a hexagonal structure of the CaCu$_5$ type  with a space group P6/mmm and
lattice parameters for x=0 --- $a$=$b$=4.8988~\AA~and $c$=3.96~\AA~(see Fig.~\ref{fig1}),
x=1 --- $a$=$b$=4.9126~\AA~and $c$=3.9901~\AA~and x=3 --- $a$=$b$=4.9351~\AA~and $c$=4.0113~\AA.\cite{Lizarraga06}
The nickel has two
inequivalent crystallographic positions Ni1(2c) (1/3,2/3,0) and Ni2(3g) (1/2,0,1/2), Tb has a position 1(a)
(0,0,0). In Fig.~\ref{fig1} Ni1 ions are represented by violet circles and located in the same ``layer'' as Tb
ions (green circles, Fig.~\ref{fig1}). The Ni2 ions (red circles, Fig.~\ref{fig1}) also form a ``layer''.

A calculations of the electronic structure of TbNi$_{5-x}$Cu$_x$ ($x=$ 0, 1, 2) were performed within the linear
muffin-tin orbitals (LMTO) method~\cite{Andersen75}. The program package TB-LMTO-ASA (tight-binding LMTO together with atomic
spheres approximation ) v.~47 was employed~\cite{Andersen84}. The subdivision of the first Brillouen zone in the reciprocal
space is (10 10 10) with 132 irreducible $\bf k$-points. The atomic sphere radius for Tb was obtained to be about 3.47
a.u., for nickel and copper ions about 2.63~a.u. The orbital basis consists of 6s, 5d and 4f muffin-tin orbitals
for Tb, and 4s, 4p and 3d for Cu and Ni. LSDA+U calculations
on top of that were done as proposed in Ref.~\onlinecite{Anis97} with Coulomb interaction parameters
for Tb(4f) states $U=$ 5.4~eV and $J=$ 0.7~eV.\cite{Knyazev06}
Also recently we showed that Coulomb interaction for Fe ions in intermetallides
is almost completely accounted within LSDA by itself.\cite{Luk09}

To introduce Cu into the TbNi$_5$ system one or two Ni ions were substituted with Cu. To mimic
disordered alloy we perform calculations for different possible positions of copper ions for
chosen doping level $x$. Assuming that each possible Cu position has the same weight we then
average obtained partial densities of Ni and Cu-3d states with following ratios for
corresponding Ni Wickhoff positions: for $x=1$ i.e. one copper ion 2(2c):3(3g) and $x=2$, two Cu ions,
1(2c):3(3g):6(2c+3g).

In our LSDA+U calculations collinear ferromagnetic order of the local magnetic moments on all
lattice sites was chosen. Moreover the terbium and nickel sublattices are taken to be
ferromagnetic with respect to each other. In this compound main magnetic moment is localized on
terbium ions while nickel ions have a small magnetization because of almost completely occupied
Ni-3d shell in accordance with Ref.~\onlinecite{Buschow}. For parent TbNi$_5$ system local
magnetic moments in units of Bohr magneton are following: on Tb(5.8), Ni1(0.27) and Ni2(0.28).
Total magnetic moment of TbNi$_5$ is 7.21~Bohr~magneton. With Ni for Cu substitution in TbNi$_5$
depending on Cu position total magnetic moment goes down about 10-15\%. 
These results agree well with experimental data of Ref.~\onlinecite{Lizarraga06}.
Local magnetic moments
on nickel ions become two to three times smaller. Local magnetic moment on terbium
insignificantly grows, while copper stays practically non magnetic.

\begin{figure}[b]
\epsfxsize=7cm \epsfbox{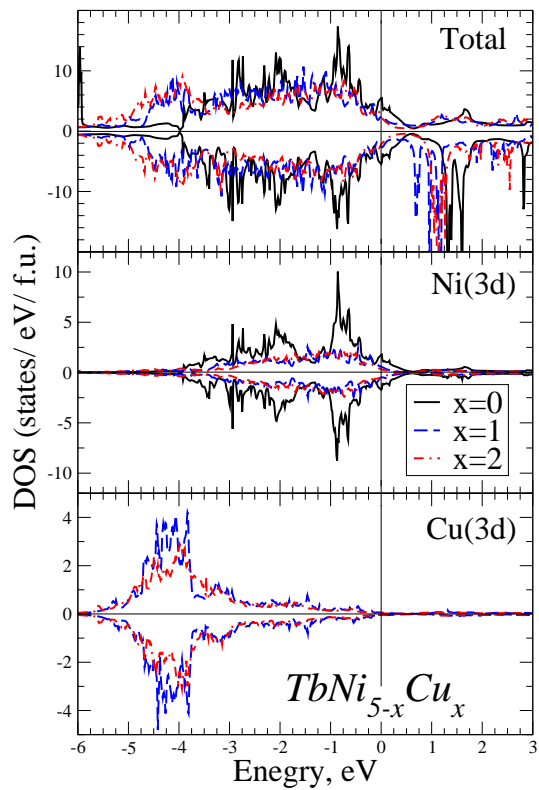}
\caption{(Color online)  LSDA+U calculated DOS for TbNi$_{5-x}$Cu$_x$. Upper panel -- total DOS;
middle panel -- Ni-3d states; lower panel -- Cu-3d states. In all graphs
solid black line represents TbNi$_5$; dashed blue lines -- averaged DOS for TbNi$_4$Cu$_1$; dash-dot red line --
averaged DOS for TbNi$_3$Cu$_2$.  Fermi level corresponds to zero.}
\label{fig2}
\end{figure}

The Fig.~\ref{fig2} shows the total (upper panel) and
partial Ni(3d) (middle panel) and Cu(3d) (lower panel) DOSes for up ($\uparrow$) and down
($\downarrow$) projections of local spin moments for the ferromagnetic TbNi$_{5-x}$Cu$_x$
(x=0,1,2) compounds. Structure of these DOS for different x is rather similar to each other
and agrees with Ref.~\onlinecite{Lizarraga06} for stoichiometric TbNi$_5$.
It is seen that the main spectral weight is located below the Fermi level.
The 3d states of Ni1 and Ni2 form the wide
bands extending from –4.0 to 0.5 eV. The 3d states of Cu1 and Cu2 bands for the compounds
with x=1,2 are located in the energy range -4$\sim$-6 eV below E$_F$. Their width is considerably
smaller with respect to Ni-3d states.
The 4f-states of Tb ion for majority spin are located below -6~eV and for minority
at about 1.5~eV.

\section{Optical conductivity results and discussion}

Calculated LSDA+U band structures presented in the previous section are used to interpret
experimental optical data~\cite{Knyazev08} for the systems under consideration. In order to
calculate theoretical interband optical conductivity $\sigmaup_{theor}$ we applied a rather
simplified technique~\cite{Berglund}  using approximation, that the direct and indirect
(involving phonons) interband transitions are equally probable. Namely we computed
$\sigmaup_{theor}$ like the integral function based on the convolution of the density of states
located both below and above the $E_F$. The electronic transitions of the d$\rightarrow$f type
are not observed in the calculation due to their low probability. This technique was
already successively applied by us to intermetallic R$_2$Fe$_{17}$ compounds~\cite{Knyazev06}.

From total LSDA+U DOS presented on upper panel of Fig.~\ref{fig2} one can observe decrease
of total DOS on the Fermi level $N(E_F)$ with Cu doping.
It happens because for x=0 $N(E_F)$ consists mostly of Ni-3d states. With doping number of Ni ions
goes down while newly introduced Cu ions do not contribute to the $N(E_F)$ (see lower panel of Fig.~\ref{fig2}).

\begin{figure}
\epsfxsize=7cm \epsfbox{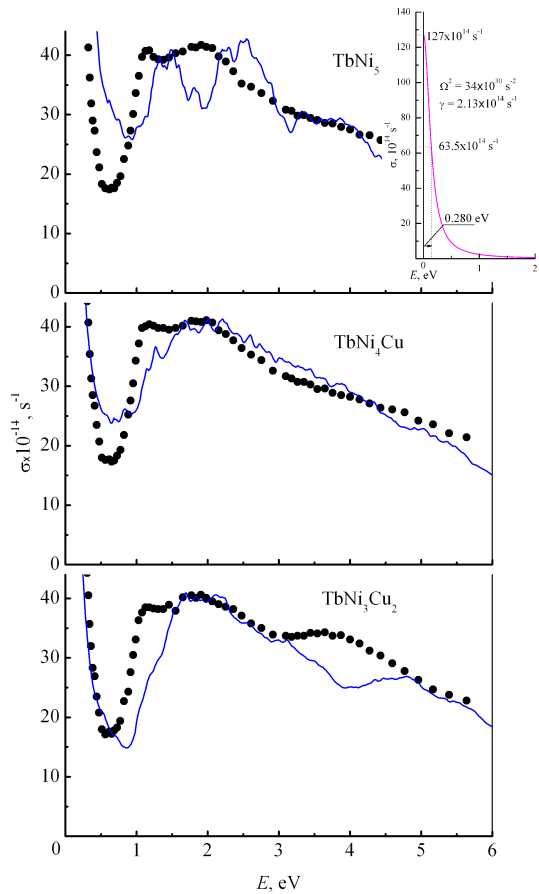}
\caption{(Color online)  Comparison of experimental optical conductivity (dots) and
theoretical one (solid line in arbitrary units) for different doping levels
x=0,1,2 (from top to bottom) for TbNi$_{5-x}$Cu$_x$. In the inset of upper panel Drude peak and
corresponding experimental parameters of the Lorentzian are shown.}
\label{fig3}
\end{figure}

Figure~\ref{fig3}  displays corresponding theoretical optical conductivity
$\sigmaup_{theor}(\omega)$ (solid lines) together with experimental
frequency dependencies of the optical conductivity $\sigmaup(\omega)$ (dots) for TbNi$_{5-x}$Cu$_x$
($x=$~0, 1, 2 from top to bottom). In the energy region below $\sim$ 0.5 eV the behavior of $\sigmaup(\omega)$ for all
compositions sharp Drude peak is determined experimentally. Corresponding height and width of Drude
peak (127$\times10^{14}$~s$^{-1}$ and 0.280~eV for $x=0$, 156$\times10^{14}$~s$^{-1}$ 
and 0.278~eV for $x=1$, 130$\times10^{14}$~s$^{-1}$ and 0.288~eV for $x=2$ ) were extracted from these 
curves and then used to draw theoretical one (see inset of Fig.~\ref{fig3}).
As follows from above mentioned static conductivity (Drude peak height) it goes down with doping.
It agrees with lowering of $N(E_F)$ in our LSDA+U calculations since conductivity is proportional
to $N(E_F)$. In its turn plasma frequency is also proportional to $N(E_F)$ and thus should
decrease with Cu doping. On the other hand substitution of Ni for Cu adds one electron to the system and
increases number of carriers. Plasma frequency squared is proportional to the number of carriers
(in the simplest Drude-like picture) and then should grow with Cu doping.
These two competing tendencies lead to the non monotonic doping behavior of plasma frequency
recently observed experimentally in Ref.~\onlinecite{Knyazev08}.

At energies above 0.7 eV the shape
of $\sigmaup(\omega)$ dispersion indicates the dominant role of interband absorption. The wide
absorption region is characterized by structures, which intensities and localizations depend on
the compound composition. The optical conductivity of the TbNi$_5$ binary alloy (upper curve)
exhibit two peaks at photon energies 1.15 and 1.9 eV respectively. In the TbNi$_4$Cu compound
the $\sigmaup(\omega)$ spectrum is almost the same; we can only note that the minimum between
these structures is less pronounced. With a further increase in the Cu content ($x=2$) a new
fairly broad absorption band arises at energies 3-4.5 eV.

Corresponding theoretical optical conductivity curves $\sigmaup_{theor}$
shows evolution upon substitution of nickel with copper similar to experimental one.
Theoretical calculations reproduce qualitatively the basic features of the
experimental spectra, namely, two-peak structure at 1-2.5 eV, associated with
the Ni~3d interband transitions.
The depth about 1.5~eV is getting closed. High energy tail above 3~eV 
becomes smoother with doping increase and for $x=2$ one can see the formation
of the ``Cu-states'' peak at about 5~eV.
Somewhat higher energy position of leading interband absorption edge (about 1 eV)
might come from uncertainties in computation of $U$ value.

\section{Conclusion}

In this work we perform {\it ab~initio} LSDA+U computations of electronic structure of
intermetallic isostructural alloys TbNi$_{5-x}$Cu$_x$ ($x=$ 0, 1, 2). Disorder effects of Ni for Cu
substitution were accounted in a combinatorial way. Corresponding averaged over all possible Cu
positions for given doping level x DOS are further used to compute theoretical optical
conductivity. Here we
is theoretically explain how experimental optical conductivity line shape changes with doping.
Analyzing  LSDA+U calculated DOS one can find that responsible for the structures of
optical conductivities at 1-2 eV electronic
states are predominantly Ni-3d states, while at $\sim$4-5~eV there are Cu-3d states.
Amount of latter ones increases upon doping giving rise to the structure at 4-5 eV in
optical conductivity. One should note also good semiquantitative agreement between theoretical
and experimental data. Finally in this work we give explanation of the experimentally observed
maximum of plasma frequency for x=1\cite{Knyazev08} as competition of lowering of $N(E_F)$ and
simultaneous growing of number of carriers with Ni for Cu substitution.

\section{Acknowledgements}

This work is partly supported by RFBR grant 08-02-00021 and was performed
within the framework of programs of fundamental research of the Russian Academy
of Sciences (RAS) ``Quantum physics of condensed matter'' (09-$\Pi$-2-1009) and 
of the Physics Division of RAS  ``Strongly correlated electrons in solid states'' 
(09-T-2-1011). IN thanks Grant of President of Russia MK-614.2009.2,
interdisciplinary UB-SB RAS project, and Russian Science Support
Foundation.

\end {document}